\documentclass[10pt]{article}
\def\prd{Phys.~Rev.~D}
\def\prl{Phys.~Rev.~Lett. }

\begin{document}

\title{Gravitational waveforms from a Lense-Thirring system}
\author{J\'anos Maj\'ar and M\'aty\'as Vas\'uth
\footnote{Electronic addresses: {\tt majar@rmki.kfki.hu, vasuth@rmki.kfki.hu}} \\ 
{\em\small KFKI Research Institute for Particle and Nuclear Physics,}\\
{\em\small Budapest 114, P.O.Box 49, H-1525 Hungary}}

\date{\today}
\maketitle

\begin{abstract}
\noindent 
The construction of ready to use templates for gravitational waves from spinning binaries 
is an important challenge in the investigation of detectable
gravitational wave signals. Here we present a method to evaluate the gravitational wave 
polarization states for inspiralling compact binaries in the extreme 
mass ratio limit. We discuss the effects caused by 
the rotation of the central massive object for eccentric orbits in the Lense-Thirring approximation
and give the formal expressions of the polarization states including higher order corrections. 
Our results are in agreement with existing calculations for the spinless and circular orbit limits.
\end{abstract}

\maketitle

\section{Introduction}

Compact objects forming binary systems are among the possible sources of gravitational radiation, whose detection is 
expected by gravitational wave observatories, i.e. LIGO \cite{LIGO}, VIRGO 
\cite{VIRGO}, GEO600 \cite{GEO600}, TAMA \cite{TAMA} and LISA \cite{LISA}. Under radiation reaction the orbit of
the binary slowly decays and the frequency and amplitude of the emitted signal increase over time. For the evaluation 
of the the noisy output of the 
detector, the method of matched filtering \cite{Matched} is employed, whose effectiveness depends on the precise 
knowledge of the gravitational waveform.

One of the commonly used approximation schemes for the computation of the dynamics and the emitted gravitational wave 
signal is the post-Newtonian (PN) expansion \cite{BDI,DJS,WW}. In this weak field approximation, 
the velocities and the gravitational potential involved are small, but there is no restriction on the mass ratio of 
the components. This approach is
considered to be a precise tool to describe the motion of an inspiralling binary up to the innermost stable circular 
orbit, which is determined with 
high accuracy in \cite{Blanchet}. Close to the merger phase the PN approximation becomes less accurate,
known as the intermediate binary black hole problem \cite{BCV,Thorne}, therefore one has to use different techniques.

The computation of the gravitational wave polarization states was carried out {\it e.g.} in \cite{LW,BIWW,BDE} and 
\cite{JS,BS,GI} for quasi-circular and elliptic orbits, respectively. For spinning binaries the evaluation of the 
wave pattern has been discussed by several authors, see {\it e.g.} \cite{WW,Kidder,ACST}. In most cases the 
gravitational waveform is expressed formally, i.e. in terms of the dynamical quantities of the motion. Up to now 
different calculations which have determined the explicit time dependence of the emitted waves analytically are 
restricted to the circular 
orbit or spinless cases.

Here we present a method to express the gravitational waveforms of a binary system using the PN approximation. 
The motion of the binary is described in the Lense-Thirring approach \cite{LT}, when we consider the geodesic motion 
around a central spinning body. We give the formal expressions of the wave polarization states $h_{+}$ and 
$h_{\times}$ up to 1.5 relative PN order beyond the leading order Newtonian expressions in the case of eccentric 
orbits. To describe the effects of rotation we focus on terms linear in the spin of the central, massive body. 
Moreover, we reproduce the circular orbit case in order to determine the influence of the 
eccentricity on the detectable gravitational waves.

Our work is based on the results of Kidder, Will and Wiseman \cite{WW,Kidder}. With the use of the general definition 
of the mass and current multipole moments they have determined the form of the transverse-traceless tensor 
$h^{ij}_{TT}$ representing metric perturbations up to 1.5 relative PN order in the case of a spinning compact binary 
in terms of the constants of the motion, 
the separation and relative velocity vectors and the vector representing the direction of the line of sight. Since 
the expressions contain these quantities in a formal way, their results can be used both in the eccentric and 
circular orbit 
cases. They have given a very useful form of the transverse-traceless tensor in the circular orbit case with the 
decomposition of the relative velocity vector into components parallel and perpendicular to the separation vector. We 
give a similar decomposition of $h^{ij}_{TT}$ in the eccentric orbit case and we use their results to check our 
formulae.

To discuss the motion of the orbiting bodies we use the results of Gergely et. al. \cite{GPV1}, where the complete 
radial and angular description of the motion is given in the Lense-Thirring approximation. The Lagrangian formalism 
was chosen as the starting point in determining the equations of motion. They have given an appropriate 
parameterization of the orbit (for further details see \cite{Param}), which can be the basis of the integration 
of the dynamics.

In order to avoid long and complicated expressions we use the comoving coordinate system, where the $x$ axis is 
aligned with the separation vector ${\bf r}$ and the $z$ axis with the newtonian angular momentum 
${\bf L_N}=\mu {\bf r}\times{\bf v}$. Here ${\bf v}$ is the relative velocity vector. We describe the motion 
in the comoving system since the transverse-traceless tensor, $h^{ij}_{TT}$, and the projection formulae are simple
in this case. 
For a convenient description of the motion we introduce a coordinate system which does not change over time, called 
invariant system. We can fix the $z$ axis of this coordinate system to the spin vector of the massive body, 
since the precession of the spin can be neglected in the Lense-Thirring approximation.

We organize our paper as follows.
In Sec. II we describe the dynamical elements of the motion in the Lense-Thirring approximation, in the invariant and 
comoving coordinate systems. For the description of the time evolution of these elements we introduce a third 
coordinate system where the $x$ and $z$ axes are aligned with the separation and angular momentum vectors, ${\bf r}$
and ${\bf L}$. The transformation between the invariant 
and comoving systems is parametrized with Euler-angles. In Sec. III we compute the explicit form of the 
vectors which describe the relative position of the binary and the detector in order to 
express the polarization states. Moreover we give the components of the spin vector in the comoving system. 
Sec. IV contains our main results, namely the splitting of the detectable gravitational 
wave signal into two polarization states $h_+$ and $h_{\times}$ and their derivation from $h^{ij}_{TT}$. 
To complete the general description of the method we show explicit contributions to the gravitational wave
signal which belong to the different polarizations, PN orders and spin effects. In Secs. V and VI we discuss 
the spinless and circular orbit limits of our calculations. 

Throughout the paper we use the $c=G=1$ convention.

\section{Description of the motion}

The components of the $h^{ij}_{TT}$ tensor have the simplest form in a comoving system, where the $x$ axis lies in 
the 
direction of the separation vector ${\bf r}$ and the $z$ axis 
in the direction of the Newtonian angular momentum ${\bf L_N}$. Since the polarization states are scalar 
quantities, we can choose this comoving system to determine their form.

In the Lense-Thirring approximation let $M$ and ${\bf S}$ denote the mass and the spin vector of the rotating central 
body and $m$ the mass of the particle ($m\ll M$) orbiting it. 

When the two bodies have comparable masses the evolution of the spin is governed by the spin precession equations 
\cite{BOC}. Since the length of the spin vector $S$ is constant the relative order of the change of the spin is
\begin{eqnarray}
\dot{{\bf S}}\sim\frac{m}{M}\epsilon\ ,
\end{eqnarray}
where $\epsilon$ is the PN parameter. When the mass ratio is negligible the precession of the spin vanishes.
In this case the $z$ and $y$ axes of the invariant system can be aligned with the direction of ${\bf S}$ and 
${\bf S}\times {\bf N}$, respectively, where ${\bf N}$ denotes the direction of the line of sight.

To describe the transformation between the comoving and the invariant systems we use Euler angles 
\cite{GPV1,GPV2}. The vector ${\bf u}$ in the comoving system can be parametrized in the 
invariant system as
\begin{eqnarray}
{\bf u}'=R_z(\Phi)R_x(\iota_N)R_z(\Psi){\bf u}\label{uvec}\ ,
\end{eqnarray}
where $\iota_N$ is the angle between ${\bf S}$ and ${\bf L_N}$, $\Phi$ represents the 
precession of the orbit (precession of ${\bf L}_N$ over ${\bf S}$), and $\Psi$ is the polar angle in
the orbital plane. In the invariant system the separation vector has the form
\begin{eqnarray}
{\bf r}=r\left(\begin{array}{c} \cos{\Phi}\cos{\Psi}-\cos{\iota_N}\sin{\Phi}\sin{\Psi} \\
\sin{\Phi}\cos{\Psi}+\cos{\iota_N}\cos{\Phi}\sin{\Psi} \\
\sin{\iota_N}\sin{\Psi}\end{array} \right) \ .
\end{eqnarray}

Using the notation of \cite{GPV1} the radial equation of motion can be written as
\begin{eqnarray}
\dot {r}^2=\frac{2E}{m}+\frac{2M}{r}-\frac{L^2}{m^2r^2}-\frac{4L_zS}{mr^3}\label{rdot}\ ,
\label{radial}
\end{eqnarray}
where the total energy $E$, the magnitude $L=\vert{\bf L}\vert$ and the $z$ component $L_z$ of the angular momentum 
${\bf L}={\bf L}_N+{\bf L}_{SO}$ are constants of the motion (the expression for the spin-orbit part ${\bf L}_{SO}$ 
of ${\bf L}$ can be found in \cite{GPV1}).

In order to get the equations for the angles, we introduce a third coordinate system called secondary system, similar 
to the comoving one. In this system the $x$ axis is aligned with the separation vector ${\bf r}$ and $z$ with the 
angular momentum ${\bf L}$. Again, every vector ${\bf u}$ of this new coordinate system in the invariant one become
\begin{eqnarray}
{\bf u'}=R_z(\Phi')R_x(\iota')R_z(\Psi'){\bf u}\ .
\end{eqnarray}
Here $\iota'$ is the angle between ${\bf S}$ and ${\bf L}$. Since the magnitude and the $z$ component of ${\bf L}$ 
($L$ and $L_z$ respectively) are constants of the motion, $\iota'$ is constant up to 1.5\,PN order. The angle 
$\Phi'$ corresponds to the precession of ${\bf L}$ over ${\bf S}$, and $\Psi'$ shows the direction of 
${\bf r}$ in the plane perpendicular to 
${\bf L}$. The equations of the motion for these angles are given as
\begin{eqnarray}\label{angular1}
\dot{\Psi'}=\frac{L}{m r^2}\ ,\quad
\dot{\Phi'}=\frac{2S}{r^3}\ .
\end{eqnarray}

The solutions of Eq.\ (\ref{angular1}) determine the equations of motion for the original angles 
(for further details see \cite{GPV1}):
\begin{eqnarray}\label{angular2}
\dot{\Psi}&=&\frac{L}{m r^2}+\frac{S\cos{\iota'}}{Lr^3}(2L-m r\dot{r}\sin{2\Psi'})\nonumber\\
\dot{\Phi}&=&\frac{2S\sin{\Psi'}}{Lr^3}(m r\dot{r}\cos{\Psi'}+2L\sin{\Psi'})\nonumber\\
\dot{\iota}_N&=&\frac{2S\sin{\iota'}\cos{\Psi'}}{Lr^3}(m r\dot{r}\cos{\Psi'}+2L\sin{\Psi'})\ .
\end{eqnarray}

For later convenience we decompose the relative velocity vector as
\begin{eqnarray}
{\bf v}=\dot{r}{\bf n}+v_{\perp}{\bf m}\label{vel}\ ,
\end{eqnarray}
where ${\bf n}={\bf r}/r$ is the direction of the $x$ axis of the comoving system and ${\bf m}$ denotes the 
direction of its $y$ axis. The relative velocity vector in this system is
\begin{eqnarray}
{\bf v}=\left(\begin{array}{c} \dot{r} \\ r(\cos{\iota_N}\dot{\Phi}+\dot{\Psi}) \\ 0 \end{array} \right)\ .
\end{eqnarray}

From the expressions of the constants of motion we can determine the magnitude and the components of the relative 
velocity vector (see \cite{GPV2}). Using the decomposition of the length of the separation vector $r=r_0+r_1S$ these
components are
\begin{eqnarray}
v=v_0-\frac{Mr_1}{r_0^2v_0}S\ , \quad
\dot{r}=(\dot{r})_0-\frac{Mr_1m^2r_0-L^2r_1+2L_zm}{m^2r_0^3(\dot{r})_0}S\ ,\quad
v_{\perp}=\sqrt{v^2-\dot{r}^2}=v_{\perp 0}+\frac{L^2r_1-2L_zm}{Lmr_0^2}S\ ,\label{vel2}
\end{eqnarray}
where we have introduced the short-hand notations
\begin{eqnarray}
v_0=\sqrt{\frac{2E}{m}+\frac{2M}{r_0}}\ , \quad
(\dot{r})_0=\sqrt{\frac{2E}{m}+\frac{2M}{r_0}-\frac{L^2}{m^2r_0^2}}\ ,\quad
v_{\perp 0}=\frac{L}{mr_0}\ .
\end{eqnarray}

\section{Basic vectors in the comoving system}

To describe the projections of $h^{ij}_{TT}$ for the calculation of the polarization states $h_+(t)$ and 
$h_{\times}(t)$, we determine the components of the orthonormal triad (${\bf N}$, ${\bf p}$, ${\bf q}$). The vector 
${\bf N}$ denotes the direction of the line of sight, ${\bf p}$ is chosen to be perpendicular to the Newtonian 
angular 
momentum (this way ${\bf p}$ lies in the direction of the node line, the intersection of the orbital plane of the 
source and the plane perpendicular to ${\bf N}$ called plane of the sky), and ${\bf q}={\bf N}\times{\bf p}$.

In the invariant system ${\bf N}=(\sin{\gamma}, 0, \cos{\gamma})$, 
where $\gamma$ is the constant angle between ${\bf N}$ and ${\bf S}$. This way the components of ${\bf N}$ in the 
comoving system are:
\begin{eqnarray}
{\bf N}=\left(\begin{array}{c} \cos{\Psi}\cos{\Phi}\sin{\gamma}-\sin{\Psi}\cos{\iota_N}\sin{\Phi}\sin{\gamma}+ 
\sin{\Psi}\sin{\iota_N}\cos{\gamma}\\ 
-\sin{\Psi}\cos{\Phi}\sin{\gamma}-\cos{\Psi}\cos{\iota_N}\sin{\Phi}\sin{\gamma}+ 
\cos{\Psi}\sin{\iota_N}\cos{\gamma} \\ \sin{\iota_N}\sin{\Phi}\sin{\gamma}+\cos{\iota_N}\cos{\gamma} \end{array} 
\right)\ .
\end{eqnarray}

There are three conditions for the vector ${\bf p}$. It is a unit vector which is perpendicular to ${\bf N}$ and 
${\bf L}_N$. From the last condition we 
get that $p_z=0$ in the comoving system. With the use of the other conditions, the form of ${\bf p}$ in the comoving 
system becomes
\begin{eqnarray}
{\bf p}=\frac{1}{N}\left(\begin{array}{c} 
\sin{\Psi}\cos{\Phi}\sin{\gamma}+\cos{\Psi}\cos{\iota_N}\sin{\Phi}\sin{\gamma}- \cos{\Psi}\sin{\iota_N}\cos{\gamma} 
\\
\cos{\Psi}\cos{\Phi}\sin{\gamma}-\sin{\Psi}\cos{\iota_N}\sin{\Phi}\sin{\gamma}+ \sin{\Psi}\sin{\iota_N}\cos{\gamma}\\ 
0 
\end{array} \right)\ ,
\end{eqnarray}
where
\begin{eqnarray}\label{N}
N=\sqrt{N_x^2+N_y^2}=\sqrt{1-N_z^2}= \sqrt{1-(\sin{\iota_N}\sin{\Phi}\sin{\gamma}+\cos{\iota_N}\cos{\gamma})^2}\ .
\end{eqnarray}

In order to determine the polarization states, we will also need the form of the spin vector in the comoving system,
\begin{eqnarray}
{\bf S}=S\left(\begin{array}{c} \sin{\Psi}\sin{\iota_N} \\ \cos{\Psi}\sin{\iota_N} \\ \cos{\iota_N} 
\end{array}\right)\ .
\end{eqnarray}

In this way we have determined all the components of the basic vectors needed to evaluate the projections of the 
transverse-traceless tensor in terms of the elements of the motion.

\section{The formal expressions of the polarization states}

The signal $h(t)$ of a laser-interferometric gravitational wave detector can be decomposed into two 
polarization states \cite{ACST} 
$h_+(t)$ and $h_{\times}(t)$
\begin{eqnarray}
h(t)=F_+(\alpha,\beta,\xi)h_+(t)+F_{\times}(\alpha,\beta,\xi)h_{\times}(t)\ .
\end{eqnarray}
The beam-pattern functions $F_+$ and $F_{\times}$ depend on the direction of the source (angles $\alpha$ and $\beta$) 
and the polarization angle ($\xi$):
\begin{eqnarray}
F_+&=&\frac{1}{2}(1+{\cos{\alpha}}^2)\cos{2\beta}\cos{2\xi}+\cos{\alpha}\sin{2\beta}\sin{2\xi}\ ,\\
F_{\times}&=&-\frac{1}{2}(1+{\cos{\alpha}}^2)\cos{2\beta}\sin{2\xi}+ \cos{\alpha}\sin{2\beta}\cos{2\xi}\ .
\end{eqnarray}
The independent polarization states of the waveform can be projected out from $h^{ij}_{TT}$ as
\begin{eqnarray}
h_+=P_{+ij}h^{ij}_{TT}\ ,\quad
h_{\times}=P_{\times ij}h^{ij}_{TT} \ .\label{h-k}
\end{eqnarray}
The components of the projectors have the form:
\begin{eqnarray}
P_{+ij}=\frac{1}{2}(p_ip_j-q_iq_j)\ ,\quad
P_{\times ij}=\frac{1}{2}(p_iq_j+q_ip_j)\ .\label{projectors}
\end{eqnarray}

In the post-Newtonian approximation up to 1.5\,PN order $h^{ij}_{TT}$ has the form \cite{Kidder}
\begin{eqnarray}
h^{ij}_{TT}=\frac{2\mu}{D}\left[Q^{ij}+P^{0.5}Q^{ij}+PQ^{ij}+PQ^{ij}_{SO}+P^{1.5}Q^{ij} 
+P^{1.5}Q^{ij}_{SO}\right]_{TT}\ ,\label{hTT}
\end{eqnarray}
where $\mu=m_1m_2/(m_1+m_2)$ is the reduced mass of the binary, and $D$ is the distance between the source and the 
observer. $Q^{ij}$ is the 
quadrupole (or Newtonian) term, $P^{0.5}Q^{ij}$, $PQ^{ij}$ and $P^{1.5}Q^{ij}$ are corrections corresponding to 
higher PN orders, $PQ^{ij}_{SO}$ and 
$P^{1.5}Q^{ij}_{SO}$ are the spin-orbit terms \cite{Kidder,WW}.

Now we evaluate the polarization states of the detectable gravitational waveform. In order to avoid long expressions, 
we use the components of 
the vectors {\bf N}, {\bf p}, {\bf q}, {\bf v} and {\bf S} in a formal way. After substituting them into Eq.\ 
(\ref{h-k}) 
and with the use of the transverse-traceless 
tensor given by Kidder \cite{Kidder} in the Lense-Thirring case we can determine the contributions to the 
polarization states. To do this we use the decomposition of the relative velocity vector (\ref{vel}), which 
suggests a natural structure to $h^{ij}_{TT}$ (see Appendix), and helps us to 
describe the projection in the comoving system. Similarly to $h^{ij}_{TT}$, the polarization 
states can be decomposed as
\begin{eqnarray}
h_{^+_{\times}}=\frac{2m}{D}\left[h_{^+_{\times}}{}^N+h_{^+_{\times}}{}^{0.5}+ 
h_{^+_{\times}}{}^1+h_{^+_{\times}}{}^{1SO}+h_{^+_{\times}}{}^{1.5}+ 
h_{^+_{\times}}{}^{1.5SO}\right]\ ,\label{waves}
\end{eqnarray}
where we have used the fact that $\mu\approx m$ in the Lense-Thirring approximation.
In this way the expressions for the "plus" polarization are:
\begin{eqnarray}
h_+^N&=&\left(\dot{r}^2-\frac{M}{r}\right)(p_x^2-q_x^2)+2v_{\perp}\dot{r}(p_xp_y-q_xq_y)+ 
v_{\perp}^2(p_y^2-q_y^2)\label{wavefirst}\\
h_+^{0.5}&=&\left[\left(\frac{2M\dot{r}}{r}-\frac{\dot{r}^3}{2}\right)N_x+ 
v_{\perp}\left(\frac{M}{2r}-\dot{r}^2\right)N_y\right](p_x^2-q_x^2)\nonumber\\ 
&+&v_{\perp}\left[\left(\frac{3M}{2r}-2\dot{r}^2\right)N_x-2v_{\perp}\dot{r}N_y\right] (p_xp_y-q_xq_y)\nonumber\\
&-&v_{\perp}^2\left[\dot{r}N_x+v_{\perp}N_y\right](p_y^2-q_y^2)\\
h_+^1&=&\frac{1}{6}\left[\left(-\frac{21\dot{r}^2M}{r}+\frac{3Mv^2}{r}+6\dot{r}^4 
+\frac{7M^2}{r^2}\right)N_x^2+4v_{\perp}\dot{r}\left(-\frac{6M}{r}+3\dot{r}^2\right)N_xN_y \right.\nonumber\\ 
&+&2v_{\perp}^2\left(3\dot{r}^2-\frac{M}{r}\right)N_y^2+ 
\left.\left(\frac{19\dot{r}^2M}{r}+3v^2\dot{r}^2-\frac{10v^2M}{r}+\frac{29M^2}{r^2}\right) 
\right](p_x^2-q_x^2)\nonumber\\
&+&\frac{1}{6}\left[6v_{\perp}\dot{r}\left(-\frac{5M}{r}+2\dot{r}^2\right)N_x^2 
+8v_{\perp}^2\left(-4\frac{M}{r}+3\dot{r}^2\right)N_xN_y+12v_{\perp}^3\dot{r}N_y^2 \right.\nonumber\\
&+&\left.6v_{\perp}\dot{r}\left(\frac{2M}{r}+v^2\right)\right](p_xp_y-q_xq_y)\nonumber\\ 
&+&\frac{1}{6}\left[2v_{\perp}^2\left(-\frac{7M}{r}+3\dot{r}^2\right)N_x^2
+12v_{\perp}^3\dot{r}N_xN_y+6v_{\perp}^4N_y^2 +v_{\perp}^2\left(-\frac{4M}{r}+3v^2\right)\right](p_y^2-q_y^2)
\\
h_+^{1SO}&=&\frac{1}{r^2}\left[({\bf qS})p_x+({\bf pS})q_x\right]\\
h_+^{1.5}&=&\left[\dot{r}\left(\frac{3\dot{r}^2M}{4r}-\frac{v^2M}{r}- 
\frac{41M^2}{12r^2}-\dot{r}^4\right)N_x^3+ v_{\perp}\left(\frac{85\dot{r}^2M}{8r}-\frac{9v^2M}{8r}-\frac{7M^2}{2r^2}- 
3\dot{r}^4\right)N_x^2N_y\right.\nonumber\\
&+& 3\dot{r}v_{\perp}^2\left(\frac{2M}{r}-\dot{r}^2\right)N_xN_y^2
+v_{\perp}^3\left(\frac{M}{4r}-\dot{r}^2\right)N_y^3+ 
\dot{r}\left(-\frac{5\dot{r}^2M}{r}+\frac{v^2M}{r}-\frac{59M^2}{12r^2}- 
\frac{1}{2}v^2\dot{r}^2\right)N_x\nonumber\\
&+&\left.v_{\perp}\left(-\frac{25\dot{r}^2M}{8r}+\frac{7v^2M}{8r}- 
\frac{13M^2}{3r^2}-\frac{1}{2}v^2\dot{r}^2\right)N_y\right](p_x^2-q_x^2)\nonumber\\
&+&v_{\perp}\left[\left(\frac{\dot{r}^2M}{4r}-\frac{7v^2M}{4r}- 
\frac{11M^2}{r^2}-2\dot{r}^4\right)N_x^3+2v_{\perp}\dot{r}\left(\frac{8M}{r}- 3\dot{r}^2\right)N_x^2N_y
\right.\nonumber\\
&+&3v_{\perp}^2\left(\frac{5M}{2r}-2\dot{r}^2\right)N_xN_y^2-2v_{\perp}^3\dot{r}N_y^3
+\left(-\frac{49\dot{r}^2M}{4r}+\frac{11v^2M}{4r}-\frac{32M^2}{3r^2}- 
v^2\dot{r}^2\right)N_x\nonumber\\
&-&\left.v_{\perp}\dot{r}\left(\frac{2M}{r}-2v^2\right)N_y\right](p_xp_y-q_xq_y)\nonumber\\
&+&v_{\perp}^2\left[-\dot{r}\left(\frac{5M}{4r}+\dot{r}^2\right)N_x^3
+v_{\perp}\left(\frac{29M}{4r}-3\dot{r}^2\right)N_x^2N_y-3v_{\perp}^2\dot{r}N_xN_y^2
-v_{\perp}^3N_y^3\right.\nonumber\\
&-&\left.\dot{r}\left(\frac{7M}{r}+\frac{v^2}{2}\right)N_x+ 
v_{\perp}\left(\frac{3M}{4r}-\frac{v^2}{2}\right)N_y\right](p_y^2-q_y^2)\\
h_+^{1.5SO}&=&\frac{2}{r^2}\left[3v_{\perp}S_z(p_x^2-q_x^2)+\dot{r}[{\bf S}\times(p_x{\bf p}-q_x{\bf 
q})]_x-2v_{\perp}[{\bf S}\times(p_x{\bf p}-q_x{\bf q})]_y\right]\ .
\end{eqnarray}
For the "cross" polarization we obtain:
\begin{eqnarray}
h_{\times}^N&=&2\left(\dot{r}^2-\frac{M}{r}\right)p_xq_x+2v_{\perp}\dot{r}(p_xq_y+q_xp_y) +2v_{\perp}^2p_yq_y\\
h_{\times}^{0.5}&=&\left[\left(\frac{4M\dot{r}}{r}-\dot{r}^3\right)N_x+ 
v_{\perp}\left(\frac{M}{r}-2\dot{r}^2\right)N_y\right]p_xq_x\nonumber\\ 
&+&v_{\perp}\left[\left(\frac{3M}{2r}-2\dot{r}^2\right)N_x-2v_{\perp}\dot{r}N_y\right] (p_xq_y+q_xp_y)\nonumber\\
&-&2v_{\perp}^2\left[\dot{r}N_x+v_{\perp}N_y\right]p_yq_y\\
h_{\times}^{1}&=&\frac{1}{3}\left[\left(-\frac{21\dot{r}^2M}{r}+\frac{3Mv^2}{r}+6\dot{r}^4 
+\frac{7M^2}{r^2}\right)N_x^2+4v_{\perp}\dot{r}\left(-\frac{6M}{r}+3\dot{r}^2\right)N_xN_y \right.\nonumber\\ 
&+&\left.2v_{\perp}^2\left(3\dot{r}^2-\frac{M}{r}\right)N_y^2+\left(\frac{19\dot{r}^2M}{r}+ 
3v^2\dot{r}^2-\frac{10v^2M}{r}+\frac{29M^2}{r^2}\right)\right]p_xq_x\nonumber\\
&+&\frac{1}{6}\left[6v_{\perp}\dot{r}\left(-\frac{5M}{r}+2\dot{r}^2\right)N_x^2 
+8v_{\perp}^2\left(-4\frac{M}{r}+3\dot{r}^2\right)N_xN_y+12v_{\perp}^3\dot{r}N_y^2 \right.\nonumber\\
&+&\left.6v_{\perp}\dot{r}\left(\frac{2M}{r}+v^2\right)\right](p_xq_y+q_xp_y)\nonumber\\ 
&+&\frac{1}{3}\left[2v_{\perp}^2\left(-\frac{7M}{r}+3\dot{r}^2\right)N_x^2
+12v_{\perp}^3\dot{r}N_xN_y+6v_{\perp}^4N_y^2 +v_{\perp}^2\left(-\frac{4M}{r}+3v^2\right)\right]p_yq_y\\
h_{\times}^{1SO}&=&\frac{1}{r^2}\left[({\bf qS})q_x-({\bf pS})p_x\right]\\
h_{\times}^{1.5}&=&\left[\dot{r}\left(\frac{3\dot{r}^2M}{2r}-\frac{2v^2M}{r}- 
\frac{41M^2}{6r^2}-2\dot{r}^4\right)N_x^3+
v_{\perp}\left(\frac{85\dot{r}^2M}{4r}-\frac{9v^2M}{4r}- \frac{7M^2}{r^2}-6\dot{r}^4
\right)N_x^2N_y\right.\nonumber\\
&+&6\dot{r}v_{\perp}^2\left(\frac{2M}{r}-\dot{r}^2\right)N_xN_y^2
+v_{\perp}^3\left(\frac{M}{2r}-2\dot{r}^2\right)N_y^3+
\dot{r}\left(-\frac{10\dot{r}^2M}{r}+\frac{2v^2M}{r}-\frac{59M^2}{6r^2}- v^2\dot{r}^2
\right)N_x\nonumber\\
&+&\left.v_{\perp}\left(-\frac{25\dot{r}^2M}{4r}+\frac{7v^2M}{4r}- 
\frac{26M^2}{3r^2}-v^2\dot{r}^2\right)N_y\right]p_xq_x\nonumber\\
&+&v_{\perp}\left[\left(\frac{\dot{r}^2M}{4r}-\frac{7v^2M}{4r}- 
\frac{11M^2}{r^2}-2\dot{r}^4\right)N_x^3+
2v_{\perp}\dot{r}\left(\frac{8M}{r}-3\dot{r}^2\right)N_x^2N_y\right.\nonumber\\
&+&3v_{\perp}^2\left(\frac{5M}{2r}-2\dot{r}^2\right)N_xN_y^2-2v_{\perp}^3\dot{r}N_y^3
+\left(-\frac{49\dot{r}^2M}{4r}+\frac{11v^2M}{4r}-\frac{32M^2}{3r^2}- 
v^2\dot{r}^2\right)N_x\nonumber\\
&-&\left.v_{\perp}\dot{r}\left(\frac{2M}{r}-2v^2\right)N_y\right](p_xq_y+q_xp_y)\nonumber\\
&+&v_{\perp}^2\left[-\dot{r}\left(\frac{5M}{2r}+2\dot{r}^2\right)N_x^3
+v_{\perp}\left(\frac{29M}{2r}-6\dot{r}^2\right)N_x^2N_y-
6v_{\perp}^2\dot{r}N_xN_y^2-2v_{\perp}^3N_y^3\right.\nonumber\\
&-&\left.\dot{r}\left(\frac{14M}{r}+v^2\right)N_x+ v_{\perp}\left(\frac{3M}{2r}-v^2\right)N_y\right]p_yq_y\\
h_{\times}^{1.5SO}&=&\frac{2}{r^2}\left[6v_{\perp}S_zp_xq_x+\dot{r}[{\bf S}\times(p_x{\bf q}+q_x{\bf 
p})]_x-2v_{\perp}[{\bf S}\times(p_x{\bf q}+q_x{\bf p})]_y\right]\ .\label{wavelast}
\end{eqnarray}

Eqs.\ (\ref{wavefirst})-(\ref{wavelast}) are the basics of our main results to determine the time (or an appropriate 
parameter) dependence of the polarization states. After solving the radial equation of motion (\ref{radial}) it can 
be used to solve the 
angular equations (\ref{angular1}) and (\ref{angular2}), and determine the components of the relative velocity 
vector. Having in hand the full description of the motion (up to 1.5\,PN order and linear in spin) one can 
express the components of the vectors ${\bf N}$, ${\bf p}$, ${\bf q}$ and ${\bf S}$ in the comoving coordinate 
system. With the substitution of these vector components into Eqs.\ (\ref{wavefirst})-(\ref{wavelast}) one can 
determine the explicit dependence of the polarization states on time (or an appropriate parameter).

To collect all the contributions due to spin effects one has to substitute the decomposition of $r$, $v$, $\dot{r}$ 
and $v_{\perp}$, as given in Eqs.\ (\ref{vel2}), into the 
expressions of the polarization states, which results
\begin{eqnarray}
h_+^{1SO}&=&\frac{1}{r_0^2}\left[({\bf qS})p_x+({\bf pS})q_x\right]\nonumber\\
h_+^{1.5SO}&=&\frac{2}{r_0^2}\left[3v_{\perp 0}S_z(p_x^2-q_x^2)+(\dot{r})_0[{\bf S}\times(p_x{\bf p}-q_x{\bf 
q})]_x-2v_{\perp 0}[{\bf S}\times(p_x{\bf p}-q_x{\bf q})]_y\right]\nonumber\\
&+&2\left[\left((\dot{r})_0(\dot{r})_1+\frac{Mr_1}{2r_0^2}\right)(p_x^2-q_x^2)+(v_{\perp 0}(\dot{r})_1+v_{\perp 
1}(\dot{r})_0)(p_xp_y-q_xq_y)+ 
v_{\perp 0}v_{\perp 1}(p_y^2-q_y^2)\right]S\\
h_{\times}^{1SO}&=&\frac{1}{r_0^2}\left[({\bf qS})q_x-({\bf pS})p_x\right]\nonumber\\
h_{\times}^{1.5SO}&=&\frac{2}{r_0^2}\left[6v_{\perp 0}S_zp_xq_x+(\dot{r})_0[{\bf S}\times(p_x{\bf q}+q_x{\bf 
p})]_x-2v_{\perp 0}[{\bf S}\times(p_x{\bf q}+q_x{\bf p})]_y\right]\nonumber\\
&+&2\left[\left(2(\dot{r})_0(\dot{r})_1+\frac{Mr_1}{r_0^2}\right)p_xq_x+(v_{\perp 0}(\dot{r})_1+v_{\perp 
1}(\dot{r})_0)(p_xq_y+q_xp_y)+2v_{\perp 0}v_{\perp 1}p_yq_y\right]S\ ,
\end{eqnarray}
where $(\dot{r})_1$ and $v_{\perp 1}$ denote the coefficients of the terms linear in spin in Eqs.\ (\ref{vel2}). We 
note that 
other terms linear in $S$ arise when we consider the evolution of the angles appearing in ${\bf p}$ and ${\bf q}$ 
which are not listed here explicitly.

\section{The $S=0$ case}

In the $S=0$ case the description of the motion can be reformulated. Since the direction of ${\bf L}_N$ do not 
change, it is useful to choose the $z$ axis of the 
invariant and the comoving systems to be the same. Formally it means that the angle $\iota_N$ is zero, and the 
separation vector in the invariant system becomes
\begin{eqnarray}
{\bf r}=r\left(\begin{array}{c} \cos{\Phi}\cos{\Psi}-\sin{\Phi}\sin{\Psi} \\
\sin{\Phi}\cos{\Psi}+\cos{\Phi}\sin{\Psi} \\ 0\end{array} \right)=r\left(\begin{array}{c} \cos{(\Phi+\Psi)} \\ 
\sin{(\Phi+\Psi)}\\ 0\end{array} \right)\ .
\end{eqnarray}
It is convenient to introduce the angle $\Upsilon=\Phi+\Psi$. The equation of motion for it is
\begin{eqnarray}
\dot{\Upsilon}=\frac{L}{m r^2}\ .
\end{eqnarray}
The radial equation of the motion becomes $\dot {r}^2=(\dot {r})^2_0$, while the relative velocity vector is
\begin{eqnarray}
{\bf v}=\left(\begin{array}{c} \dot{r} \\ r\dot{\Upsilon} \\ 0 \end{array} \right)=\left(\begin{array}{c} \dot{r} \\ 
\frac{L}{m r} \\ 0 \end{array} \right)\ .
\end{eqnarray}
The $x$ and $y$ axes of the invariant system are again determined by the vector ${\bf N}$. Since $N=\sin{\gamma}$ 
in the comoving system, the vectors ${\bf N}$, ${\bf p}$ and ${\bf q}$ become:
\begin{eqnarray}
{\bf N}=\left(\begin{array}{c} \cos{\Upsilon}\sin{\gamma} \\ -\sin{\Upsilon}\sin{\gamma} \\ \cos{\gamma} \end{array} 
\right)\ ,\qquad
{\bf p}=\left(\begin{array}{c} \sin{\Upsilon} \\ \cos{\Upsilon} \\ 0 \end{array} \right)\ ,\qquad
{\bf q}=\left(\begin{array}{c} -\cos{\Upsilon}\cos{\gamma} \\ \sin{\Upsilon}\cos{\gamma} \\ \sin{\gamma} \end{array} 
\right)\ .
\end{eqnarray}
Although some of the formal expressions do not change in the spinless limit as compared to the general case, one has 
to substitute $S=0$ into the different expressions. Also the spin-orbit terms in $h_+$ and 
$h_{\times}$ automatically vanish.

\section{The circular orbit case}

A general decomposition of the relative velocity vector is used to describe the circular orbit case:
\begin{eqnarray}
{\bf v}=\dot{r}{\bf n}+r\omega{\bf m}\label{vcirc1}\ .
\end{eqnarray}
Using Euler angles the frequency $\omega$ has the form
\begin{eqnarray}
\omega=\cos{\iota_N}\dot{\Phi}+\dot{\Psi}\ .
\end{eqnarray}

The circular orbit case corresponds to $\dot{r}=0$ and $\dot{\omega}=0$. If we use these rules and the 
method given in \cite{Kidder} for the length of the relative velocity vector we get the following formula
\begin{eqnarray}
v^2=v_{\perp}^2=r^2\omega^2=\frac{M}{r}\left[1-3\left(\frac{M}{r}\right)-\frac{2}{M^2}({\bf n}\times{\bf m}){\bf 
S}\left(\frac{M}{r}\right)^{3/2}+6\left(\frac{M}{r}\right)^2\right]\ .\label{vcirc}
\end{eqnarray}

Substituting Eq.\ (\ref{vcirc}) for the components of the relative velocity vector into 
Eqs.\ (\ref{wavefirst})-(\ref{wavelast}), we 
get the polarization states of the detectable gravitational waves in the circular orbit limit. Again, 
we use the decomposition of $h_+$ and $h_{\times}$ as given in Eq.\ (\ref{waves}). The individual terms which 
correspond to different PN orders and spin effects are
\begin{eqnarray}
h_+^N&=&-\left(\frac{M}{r}\right)\left[(p_x^2-p_y^2)-(q_x^2-q_y^2)\right]\label{cicrfirst}\\
h_+^{0.5}&=&\left(\frac{M}{r}\right)^{3/2}\left[\frac{1}{2}N_y(p_x^2-q_x^2)+3N_x(p_xp
_y-q_xq_y)-N_y(p_y^2-q_y^2)\right]\\
h_+^1&=&\frac{1}{6}\left(\frac{M}{r}\right)^2\left[\left(10N_x^2-2N_y^2+19\right)(p_x^2-q_x^2) 
-32N_xN_y(p_xp_y-q_xq_y)\right.\nonumber\\
&+&\left.\left(-14N_x^2+6N_y^2-19\right)(p_y^2-q_y^2)\right]\\
h_+^{1SO}&=&-\left(\frac{M}{r}\right)^2\frac{1}{M^2}\left[({\bf qS})p_x+({\bf pS})q_x\right]\\
h_+^{1.5}&=&\left(\frac{M}{r}\right)^{5/2}\left[\left(-\frac{37}{8}N_x^2+\frac{1}{4}N_y^2- 
\frac{101}{24}\right)N_y(p_x^2-q_x^2)+\left(-\frac{65}{12}N_x^2+\frac{15}{2}N_y^2- 
\frac{149}{12}\right)N_x(p_xp_y-q_xq_y)\right.\nonumber\\
&+&\left.\left(\frac{29}{4}N_x^2-N_y^2+\frac{19}{4}\right)N_y(p_y^2-q_y^2)\right]\\
h_+^{1.5SO}&=&\left(\frac{M}{r}\right)^{5/2}\frac{2}{M^2}\left[\left(2(p_x^2-q_x^2)- 
(p_y^2-q_y^2)\right)S_z-q_xq_zS_x\right]
\end{eqnarray}
and
\begin{eqnarray}
h_{\times}^N&=&-2\left(\frac{M}{r}\right)\left[p_xq_x-p_yq_y\right]\\
h_{\times}^{0.5}&=&\left(\frac{M}{r}\right)^{3/2}\left[N_yp_xq_x+3N_x(p_xq_y+q_xp_y)- 2N_yp_yq_y\right]\\
h_{\times}^1&=&\frac{1}{3}\left(\frac{M}{r}\right)^2\left[\left(10N_x^2-2N_y^2+19\right)p_xq_x- 
16N_xN_y(p_xq_y+q_xp_y)\right.\nonumber\\
&+&\left.\left(-14N_x^2+6N_y^2-19\right)p_yq_y\right]\\
h_{\times}^{1SO}&=&-\left(\frac{M}{r}\right)^2\frac{1}{M^2}\left[({\bf qS})q_x-({\bf pS})p_x\right]\\
h_{\times}^{1.5}&=&\left(\frac{M}{r}\right)^{5/2}\left[\left(-\frac{37}{4}N_x^2+\frac{1}{2}N_y^2- 
\frac{101}{12}\right)N_yp_xq_x+\left(-\frac{65}{12}N_x^2+\frac{15}{2}N_y^2-\frac{149}{12}\right) 
N_x(p_xq_y+q_xp_y)\right.\nonumber\\
&+&\left.\left(\frac{29}{2}N_x^2-2N_y^2+\frac{19}{2}\right)N_yp_yq_y\right]\\
h_{\times}^{1.5SO}&=&\left(\frac{M}{r}\right)^{5/2}\frac{4}{M^2}\left[S_z(p_xq_x-p_yq_y)+ p_xq_zS_x\right]\ .
\label{cicrlast}
\end{eqnarray}

In the circular orbit case one can make the time dependence explicit since the length of the separation vector is 
constant. The equations of motion can be solved and the polarization states can be given in terms of time 
following the method described in the general case. 

Eqs.\ (\ref{cicrfirst})-(\ref{cicrlast}) are found to be in agreement with the results given in \cite{Kidder}. 
To see this one has to reexpress Eqs.\ (4.8) and (4.9) of \cite{Kidder}, using the Lense-Thirring approximation 
and evaluate the projections given in (\ref{h-k}).

\section{The relation to existing results}

The properties of gravitational waves generated by compact binaries have been studied extensively. Typical 
parameters of the binary, masses, eccentricity and spins, etc. are included in these descriptions. 
To find the relation between our results and the ones known in the literature, we compare the method presented 
here with integrated expressions of the waveforms depending explicitly on time or a chosen parameter. 
In some cases the metric perturbations $h^{ij}_{TT}$ are given formally in terms of the separation vector, the 
relative velocity and spin vectors and ${\bf N}$ \cite{Kidder,WW}. 
To express the polarization states $h_+$ and $h_{\times}$ one has to introduce suitable coordinate systems and
integrate the equations of the motion for the angle variables. For this purpose we have used Euler angles and the 
results of \cite{GPV1}.

Gopakumar and Iyer have obtained the polarization states for eccentric binaries up to 2\,PN order. To the required
accuracy they have included all radiative moments \cite{GI1} and employed the 2\,PN accurate generalized
quasi-Keplerian parameterization for non-spinning bodies \cite{GI}. Here we have used the 
1.5\,PN expression of $h^{ij}_{TT}$ and the lowest order spin contributions to the motion of the binary 
in the Lense-Thirring approximation. Higher order corrections to the waveform which arise from an appropriate 
parameterization of the motion can be investigated by substituting the result of the PN expansion of the motion 
into the quadrupole formula \cite{gop,gk}.

The effects caused by the rotation of the bodies on the waveform are studied {\it e.g.} for circular orbits 
\cite{ACST,Vecchio} or in the case of binaries with equal masses and one spinning body \cite{gop}.

Most of the results containing explicit time dependent contributions to the waveforms are valid in the circular 
orbit limit. To check our formulae in this case we have used the results of Kidder \cite{Kidder}, where the 
$h^{ij}_{TT}$ tensor is given after the decomposition of the velocity vector, Eq.\ (\ref{vcirc1}). 
As an other check we have 
integrated the equations of the motion in the circular orbit limit. With the use of the method presented above we 
have determined the lowest order contributions to the detectable waveforms. Our expressions reproduce the well known 
result of the quadrupole formalism, namely that the frequency of gravitational waves is twice the orbital frequency. 
It is worth mentioning that the eccentricity of the orbit and the lowest order PN perturbation give rise to higher 
harmonics.

\section{Conclusions and remarks}

We have presented a method to obtain the detectable gravitational wave signals generated by a spinning compact binary 
system in the Lense-Thirring approximation. This is useful for the investigation of the effects caused by the 
rotation of the components in the case of eccentric orbits. We have introduced a comoving coordinate system and shown 
how to express the wave polarizations in this system. Our formal expressions are determined in terms of the 
components of the separation vector, the relative velocity, the spin and the orthonormal triad (${\bf N}$, ${\bf p}$, 
${\bf q}$). We have calculated these vector components in terms of the angles describing the motion and the relative 
position of the source and the detector.

We have clarified all the steps needed to determine the explicit parameter dependence of $h_+$ and 
$h_{\times}$ up to 1.5\,PN relative order. To do this one can use the generalized true anomaly parameterization of 
the 
radial motion \cite{Param}, which is a natural parameterization of an eccentric orbit in the case of a spinning 
binary.

We have investigated the main characteristics of the spinless and the circular orbit limits of our general method. 
We have explored the effects of the spin and the eccentricity of the orbit. The results in the 
circular orbit limit can help one to describe the emitted waves near the innermost stable circular orbit and provide  
initial information for the methods investigating the coalescence of the binary. 

After calculating the time dependence of the detectable gravitational waves following the 
method given above one can construct an experimental receipt to determine the main properties 
of a compact binary system with the detection of the waves it emits. Although recent 
experiments cannot detect waves emitted by a Lense-Thirring system, the method can be useful 
starting point to investigate the description of the gravitational wave signals in more general 
cases analytically.

\section*{Acknowledgments}
The authors thank P\'eter Forg\'acs and L\'aszl\'o \'A. Gergely for discussions and 
helpful remarks on the manuscript. This work was supported by OTKA grants no. 
TS044665, T046939 and F049429.

\appendix

\section{The $h^{ij}_{TT}$ tensor in the Lense-Thirring case}

If we use the Lense-Thirring approximation, and decompose the relative velocity vector according to Eq.\ (\ref{vel}), 
the various contributions of the $h^{ij}_{TT}$ tensor, given in \cite{Kidder}, representing different PN orders and 
effects become
\begin{eqnarray}
Q^{ij}&=&2\left(\left[\dot{r}^2-\frac{M}{r}\right]n^in^j+2v_{\perp}\dot{r}m^{(i}n^{j)}+ v_{\perp}^2m^im^j\right)\\
P^{0.5}Q^{ij}&=&\left(\dot{r}\left[\frac{4M}{r}-\dot{r}^2\right]({\bf 
nN})+v_{\perp}\left[\frac{M}{r}-2\dot{r}^2\right]({\bf mN})\right)n^in^j\nonumber\\ 
&+&v_{\perp}\left(2\left[\frac{3M}{r}-2\dot{r}^2\right]({\bf nN})-4v_{\perp}\dot{r}
({\bf mN})\right)m^{(i}n^{j)}\nonumber\\
&-&2v_{\perp}^2\left(\dot{r}({\bf nN})+v_{\perp}({\bf mN})\right)m^im^j\\
PQ^{ij}&=&\frac{1}{3}\left(\left[-\frac{21\dot{r}^2M}{r}+\frac{3Mv^2}{r}+6\dot{r}^4 +\frac{7M^2}{r^2}\right]({\bf 
nN})^2+ 
4v_{\perp}\dot{r}\left[-\frac{6M}{r}+3\dot{r}^2\right]({\bf nN})({\bf mN})\right.\nonumber\\
&+&\left.2v_{\perp}^2\left[3\dot{r}^2-\frac{M}{r}\right]({\bf 
mN})^2+\left[\frac{19\dot{r}^2M}{r}+3v^2\dot{r}^2-\frac{10v^2M}{r}+\frac{29M^2}{r^2}
\right] 
\right) n^in^j\nonumber\\
&+&\frac{v_{\perp}}{3}\left(6\dot{r}\left[-\frac{5M}{r}+2\dot{r}^2\right]({\bf nN})^2 
+8v_{\perp}\left[-4\frac{M}{r}+3\dot{r}^2\right]({\bf nN})({\bf 
mN})+12v_{\perp}^2\dot{r}({\bf mN})^2\right.\nonumber\\
&+&\left.6\dot{r}\left[\frac{2M}{r}+v^2\right]\right)m^{(i}n^{j)}\nonumber\\ 
&+&\frac{v_{\perp}^2}{3}\left(2\left[-\frac{7M}{r}+3\dot{r}^2\right]({\bf 
nN})^2
+12v_{\perp}\dot{r}({\bf nN})({\bf mN})+6v_{\perp}^2({\bf mN})^2+\left[-\frac{4M}{r}+3v^2\right]\right)m^im^j\\
PQ^{ij}_{SO}&=&-\frac{2}{r^2}[{\bf S}\times{\bf N}]^{(i}{\bf n}^{j)}\\
P^{1.5}Q^{ij}&=&\left(\dot{r}\left[\frac{10\dot{r}^2M}{r}-\frac{2v^2M}{r}- \frac{41M^2}{6r^2}-2\dot{r}^4\right]({\bf 
nN})^3+v_{\perp}\left[\frac{85\dot{r}^2M}{4r}-\frac{9v^2M}{4r}- \frac{7M^2}{r^2}-6\dot{r}^4\right]({\bf nN})^2({\bf 
mN})\right.\nonumber\\
&+&6\dot{r}v_{\perp}^2\left[\frac{2M}{r}-\dot{r}^2\right]({\bf nN})({\bf mN})^2
+v_{\perp}^3\left[\frac{M}{2r}-2\dot{r}^2\right]({\bf 
mN})^3+\dot{r}\left[-\frac{10\dot{r}^2M}{r}+\frac{2v^2M}{r}-\frac{59M^2}{6r^2}\right.
\nonumber\\
&-&\left.\left.v^2\dot{r}^2\right]({\bf nN})+\left[-\frac{25v_{\perp}\dot{r}^2M}{4r}+\frac{7v^2v_{\perp}M}{4r}- 
\frac{26v_{\perp}M^2}{3r^2}-v^2v_{\perp}\dot{r}^2\right]({\bf mN})\right)n^in^j\nonumber\\
&+&v_{\perp}\left(\left[\frac{35\dot{r}^2M}{2r}-\frac{7v^2M}{2r}- \frac{22M^2}{r^2}-4\dot{r}^4\right]({\bf 
nN})^3+v_{\perp}\dot{r}\left[\frac{32M}{r}-12\dot{r}^2\right]({\bf nN})^2({\bf mN})\right.\nonumber\\
&+&3v_{\perp}^2\left[\frac{5M}{r}-4\dot{r}^2\right]({\bf nN})({\bf mN})^2-4v_{\perp}^3\dot{r}({\bf mN})^3
+\left[-\frac{49\dot{r}^2M}{2r}+\frac{11v^2M}{2r}-\frac{64M^2}{3r^2}-2v^2\dot{r}^2\right]({\bf nN})\nonumber\\
&-&\left.4v_{\perp}\dot{r}\left[\frac{M}{r}-v^2\right]({\bf mN})\right)m^{(i}n^{j)}\nonumber\\
&+&v_{\perp}^2\left(2\dot{r}\left[\frac{3M}{r}-\dot{r}^2\right]({\bf nN})^3
+v_{\perp}\left[\frac{29M}{2r}-6\dot{r}^2\right]({\bf nN})^2({\bf mN})-6v_{\perp}^2\dot{r}({\bf nN})({\bf mN})^2
-2v_{\perp}^3({\bf mN})^3\right.\nonumber\\
&-&\left.\dot{r}\left[\frac{14M}{r}+v^2\right]({\bf nN})+v_{\perp}\left[\frac{3M}{2r}-v^2\right]({\bf 
mN})\right)m^im^j\\
P^{1.5}Q^{ij}_{SO}&=&\frac{4}{r^2}\left(3v_{\perp}[{\bf n}\times{\bf m}]{\bf S}n^in^j+\dot{r}n^{(i}[{\bf n}\times{\bf 
S}]^{j)}-2v_{\perp}n^{(i}
[{\bf m}\times{\bf S}]^{j)}\right)\ .
\end{eqnarray}

\end{document}